\documentclass[prl,twocolumn,showpacs,hyperref,balancelastpage]{revtex4}%
\usepackage{epsfig}
\usepackage{color}
\usepackage{ulem}
\usepackage{amsmath}
\usepackage{amsfonts}
\usepackage{amssymb}
\usepackage{graphicx}%
\usepackage{pdfpages}
\providecommand{\U}[1]{\protect\rule{.1in}{.1in}}

\newcommand{\be}{\begin{equation}}
\newcommand{\ee}{\end{equation}}
\newcommand{\bea}{\begin{eqnarray}}
\newcommand{\eea}{\end{eqnarray}}

\begin{document}
\title{Fractal energy spectrum of a polariton gas in a Fibonacci quasi-periodic potential}
\author{D. Tanese$^1$, E. Gurevich$^2$, F. Baboux$^1$, T. Jacqmin$^1$, A. Lema\^{\i}tre$^1$, E. Galopin$^1$, I. Sagnes$^1$, A. Amo$^1$, J. Bloch$^1$, E. Akkermans$^2$}
\affiliation{$^1$Laboratoire de Photonique et de Nanostructures, LPN/CNRS, Route de Nozay, 91460 Marcoussis, France}
\affiliation{$^2$Department of Physics, Technion Israel Institute of Technology, Haifa 32000, Israel}

\begin{abstract}

We report on the study of a polariton gas confined in a quasi-periodic one dimensional cavity, described by a Fibonacci sequence. Imaging the  polariton modes both in real and reciprocal space, we observe features characteristic of their fractal energy spectrum such as the opening of mini-gaps obeying the gap labeling theorem and log-periodic oscillations of the integrated density of states. These observations are accurately reproduced solving an effective 1D Schr\"{o}dinger equation, illustrating the potential of cavity polaritons  as a quantum simulator in complex topological geometries.

\end{abstract}


\date{\today}
\maketitle

Free quantum particles or waves propagating in a spatially varying potential present modifications of their spectral density, which depend on the symmetry of this potential.  The richness of spectral distributions in constrained geometries has long been recognized.
The case of a periodic potential described by means of the Bloch theorem is a significant example. The notion of spectral distribution has been deepened in the wake of quasi-crystals discovery and it led to a classification of energy spectra into absolutely continuous, pure point and singular continuous spectral distributions \cite{singcontgeneral}. The latter class proved to be surprisingly rich and it encompasses a broad range of potentials, such as quasi-periodic potentials which have been thoroughly studied \cite{damanik,vardeny}.

An interesting quasi-periodic potential can be designed using a Fibonacci sequence. The corresponding singular continuous energy spectrum has a fractal structure of the Cantor set type \cite{Kohmoto-87,gelerman,QP-Review03,Kohmoto-87-Wurtz-88}, and it displays self-similarity {\it i.e.}, a symmetry under a discrete scaling transformation. Denoting $\rho (\varepsilon)$ the relevant density of states (DOS) in $\varepsilon$ (either energy or frequency), a  \textit{discrete scaling symmetry} about a particular value $\varepsilon_u$ is expressed by the
property%
\begin{equation}
\mu(\varepsilon_{u}+\Delta \varepsilon)-\mu(\varepsilon_{u})={\frac{{\mu}\left(  \varepsilon_{u}+\beta \Delta \varepsilon\right)
{-\mu\left(  \varepsilon_{u}\right)  }}{\alpha},} \label{scaling}%
\end{equation}
where $\mu\left(  \varepsilon\right)  =\int_{-\infty}^{\varepsilon}\rho\left(  \varepsilon^{\prime}\right)
d\varepsilon^{\prime}$ is the integrated density of states (IDOS), or density measure, and $\alpha$ and $\beta$
are scaling parameters which usually, depend on $\varepsilon_{u}$.
Defining a shifted IDOS by $\mathcal{N}_{\varepsilon_{u}}\left(  \varepsilon\right)  \equiv$ $\mu
(\varepsilon)-\mu\left(  \varepsilon_{u}\right)  $, the general solution of (\ref{scaling}) can be
written as \cite{reviewfractals}%
\begin{equation}
{\mathcal{N}_{\varepsilon_{u}}\left(  \varepsilon\right)  =|\varepsilon-\varepsilon_{u}|^{\gamma}\,\mathcal{F}\left(
{\frac{\ln|\varepsilon-\varepsilon_{u}|}{\ln \beta}}\right)  ,} \label{Scaling func}%
\end{equation}
where $\gamma=\frac{\ln \alpha}{{\ln \beta}}$ is the local ($\varepsilon_{u}$-dependent) scaling
exponent and $\mathcal{F}(z)$ is a periodic function of period unity, whose
(non-universal) form depends on the problem at hand. Generally, the exponent
$\gamma$ takes values between zero and unity, so that the density $\rho\left(
\varepsilon\right)  $ is a singular function. Such scaling properties of a fractal spectrum are expected to modify the behavior of
physical quantities \cite{reviewfractals}. Recently studied examples include thermodynamic properties of photons \cite{ADT2},
random walks \cite{ADT3}, quantum diffusion of
wave packets \cite{Guarneri} and spontaneous emission triggered by a fractal vacuum \cite{eagurevich}. The diffusion of a wave packet in a quasi-periodic medium is predicted to be neither diffusive, nor ballistic but to present a behavior characterized by non-universal exponents and a log-periodic modulation of its time dynamics. Experimental demonstration of these specific properties of quasi-periodic structures is still missing as yet. We propose to use cavity polaritons to evidence such a fractal behavior.

Cavity polaritons are quasi-particles arising from the strong coupling between the optical mode of an optical cavity and excitons confined in quantum wells \cite{weisbuch92}. They have appeared recently as a promising system to realize quantum simulators \cite{LaiNature2007,CarusottoRMP}. Engineering of the potential landscape is possible
and allows implementing a large variety of physical  situations such as 1D  \cite{LaiNature2007,CerdaPRL,Tanese} and 2D periodic potentials \cite{NaYongNaturePhysics2011,CerdaDots} with the generation of gap solitons \cite{Tanese,Cerda Gap solitons}, non-linear resonant tunneling devices\cite{Nguyen},  or triangular  \cite{NaYongNJPhysics} and honeycomb \cite{yamamotoHoneycomb,TJacqmin} lattices, which  enables the exploration of graphene physics. Polaritons offer experimental possibilities not available in 1D or 2D photonic quasi-crystals such as direct time- and
energy-resolved measurements of the excitations in both space and momentum
domains. Thus, one can directly visualize individual
eigenmodes, and the dynamics of wave packets.

In this letter, we use this well-controlled system to investigate both theoretically and experimentally the spectral properties of a polariton gas in a quasi-periodic potential. To do so, we have sculpted the lateral profile of a quasi-1D cavity in the shape of a Fibonacci sequence. Using non resonant excitation in the low density regime, we probe the modes both in real and reciprocal space.
We observe a quantitative agreement between experiments and the calculated modes and density of states. In particular, we evidence  features of a fractal energy spectrum, namely gaps densely distributed and an integrated density of states (IDOS) reflecting the existence of a discrete scaling symmetry as expressed by (\ref{Scaling func}).

\begin{figure}[htb]
\begin{center}
\includegraphics[width=8.5cm]{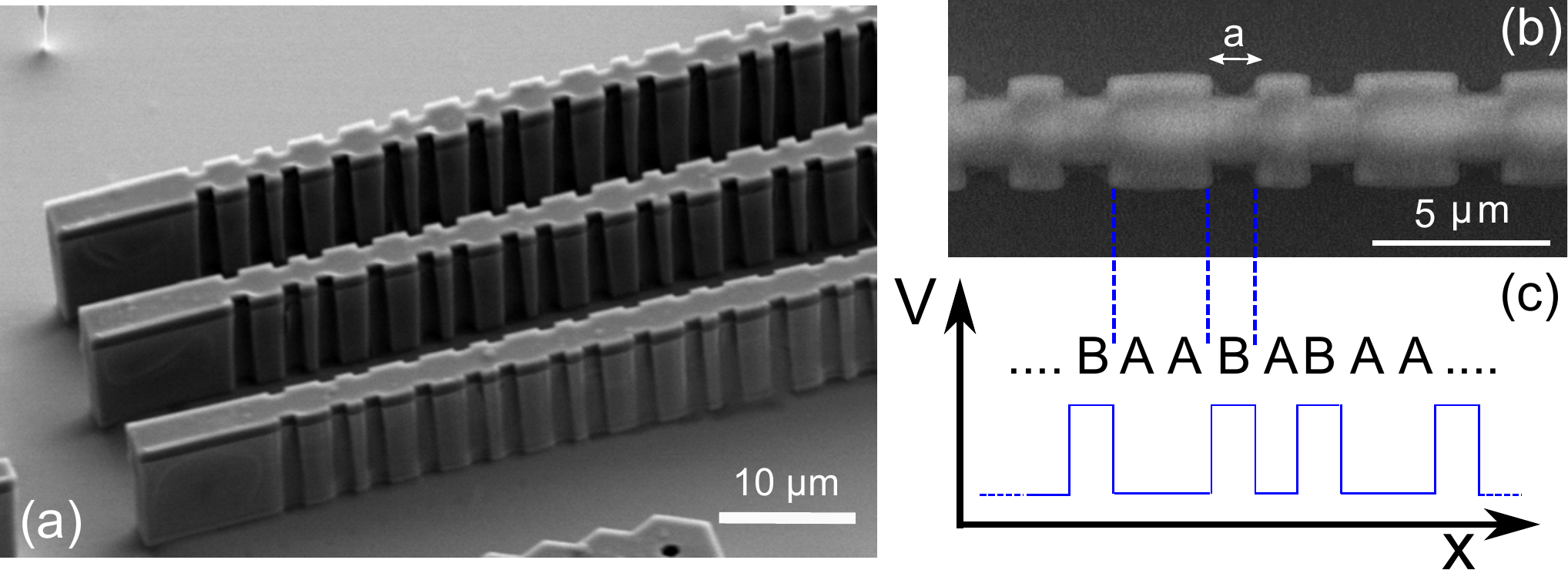}
\end{center}
\caption{\label{fig1}{(Color online) (a) Scanning electron microscopy image of an array of modulated wires. (b) Zoom on a particular wire, showing the shape of the A and B letters. (c) Schematic of the nominal potential corresponding to the lateral shaping of the wire cavity. }}
\end{figure}

In our sample, cavity polaritons are confined within narrow strips (wire cavities), whose width is modulated quasi-periodically. These wires are fabricated processing a planar high quality factor ($Q \sim 72000$) microcavity  grown by molecular beam epitaxy. It  consists in a $ \lambda /2$ $Ga_{0.05}Al_{0.95}As$ layer surrounded by two $Ga_{0.8}Al_{0.2}As/Ga_{0.05}Al_{0.95}As$ Bragg mirrors with 28 and 40 pairs in the top/bottom mirrors respectively. 12 GaAs quantum wells of width 7 nm are inserted in the structure resulting in a $15 \,meV$ Rabi splitting.  $200\,\mu m$ long wires with lateral dimension modulated quasi-periodically are designed using electron beam lithography and dry etching  (Figs.
\ref{fig1}(a-b)).  The modulation
consists in two wire sections ("letters") $A$ and $B$ of same length $a$ but different widths $w_{A}$ and $w_{B}$ respectively (Fig.\ref{fig1}(b)). The modulation of the wire width induces an effective 1D
potential for the longitudinal motion of polaritons, as discussed in the sequel. The letters are arranged according
to the Fibonacci sequence \cite{Kohmoto-87} using the recursion,
\begin{equation}
S_{j\geq3}=\left[  S_{j-2}S_{j-1}\right]  ,\text{ and }S_{1}=B,\text{ }%
S_{2}=A,\label{Fib Recurs}%
\end{equation}
where $\left[  S_{j-2}S_{j-1}\right] $ means concatenation of two
sub-sequences $S_{j-2}$ and $S_{j-1}$. The number of letters (length) of a sequence $S_j$ is given by the Fibonacci number $F_j$, such that $F_{j+1} = F_j + F_{j-1}$. The ratio $F_{j+1} / F_j$ tends to the golden mean $\sigma = (1 + \sqrt{5}) /2 \simeq 1.62$ in the limit $j \rightarrow \infty$, while the corresponding sequence $S_\infty$ becomes rigorously quasi-periodic and invariant, {\it i.e.} self-similar, under the iteration transformation (\ref{Fib Recurs}).  Our sample corresponds to
$S_{13}$ counting  $233$ letters  with $a=0.8\,\mu m$,  $w_{A}=3.5\,\mu m$ and $w_{B}=1.86\,\mu m$. 
To study the polariton modes in these quasi-periodic wires, we perform low temperature (10 K) micro-photoluminescence experiments. Single wires are excited non-resonantly using a cw monomode laser tuned typically $100\, meV$ above the polariton resonances. The excitation spot extends over a $80\,\mu m$-long region along  the wire. The sample emission is collected with a 0.65 numerical aperture objective and focused on the entrance slit (parallel to the wire) of a spectrometer coupled to a CCD camera. Imaging of the sample surface (resp. the Fourier plane of the collection objective) allows studying the spectrally resolved polariton modes in real (resp. reciprocal) space. Excitation power is kept low enough to stay below condensation threshold and obtain a nearly homogeneous population of the lower energy polariton states.

\begin{figure}[htb]
\begin{center}
\includegraphics[width=8.5cm]{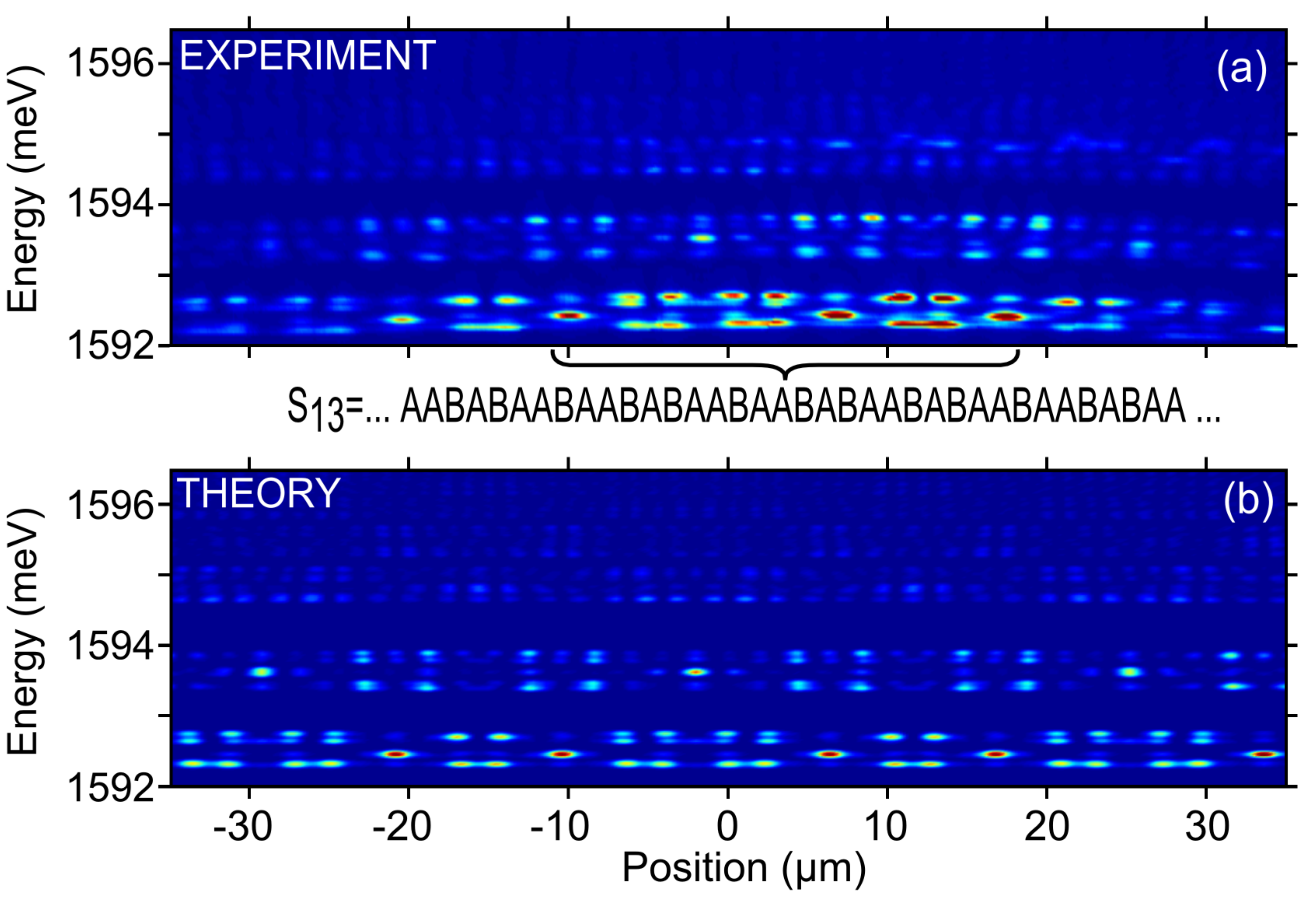}
\end{center}
\caption{\label{fig2}{(Color online) (a) Spectrally and spatially resolved emission measured on a single modulated wire (the linear polarization parallel to the wire is selected). Bottom of the figure:  letter sequence corresponding to a part of the whole $S_{13}$ potential sequence. (b) Calculated polariton Fibonacci modes as a function of energy and real space coordinate.}}
\end{figure}

Fig.\ref{fig2}.a displays the spatially and spectrally resolved emission measured on a single modulated wire cavity for an exciton-photon detuning around $-8\,meV$ (defined as the energy difference between the cavity mode at normal incidence and the exciton resonance). Several polariton modes are imaged. They present complex patterns of bright spots distributed all over the region of the wire under investigation. To understand the nature of these modes and properties of their spectral density, we have calculated the polariton eigenstates in such quasi-periodic structures.

In our model whose details are given in the supplement \cite{supplement}, we describe the confined photon modes using a 2D scalar wave equation with vanishing boundary conditions on the boundary of the wire, considered as an axially symmetric strip where the longitudinal coordinate $x\in\lbrack0,L]$ ($L$ being the length of the wire), and the transverse coordinate $-\frac{w\left(
x\right)  }{2}\leq y\leq\frac{w\left(  x\right)  }{2}$. Here,
$w\left(  x\right)  >0$  accounts for the $x$-dependent width of the wire (Fig.\ref{fig1}.c), {\it i.e.}  a quasi-periodic sequence of segments of width $w_A$ and $w_B$, as
defined in (\ref{Fib Recurs}). In the supplement \cite{supplement}, we show how to map this 2D problem onto a 1D Schr\"odinger equation with the  effective potential: 
\be
V(x) = \frac{\pi^{2}}%
{w^{2}\left(  x\right)  }+\frac{\pi^{2}+3}{12}\left(  \frac{w^{\prime}\left(
x\right)  }{w\left(  x\right)  }\right)  ^{2} \, \, .
\label{Effective 1D1}%
\end{equation}
The first term of $V(x)$ is the usual adiabatic approximation. The second term accounts for the sharpness of the steps. It is not perturbative, and it cannot be neglected (see supplement \cite{supplement}).
As clearly visible on Fig.\ref{fig1}, the strip shape is not perfectly abrupt but presents some smoothness in the width
variation introduced by the actual etching process. The smoothness scale is used as a fitting parameter in the calculations. The eigenfunctions $\phi_{q}\left(  x\right)  $ and eigenenergies $E_{C,q}$ are obtained numerically. To calculate the polariton modes, we consider the radiative coupling between excitons with a flat dispersion to the photon
modes we have obtained in our simulations. Since the coupling is diagonal in the index $q$, the resulting polariton eigenfunctions and photons have the same spatial behavior.
Fig. \ref{fig2}.b shows the polariton modes thus obtained numerically. Since experimentally we cannot resolve states which are separated by less than the polariton linewidth, we have averaged the intensity over eigenmodes
close in energy. Thus, what appears in Fig.\ref{fig2}.b as bright intensity spots at different energies are actually bands separated by gaps. Clearly the calculation reproduces very accurately the spatial structure of the  polariton modes observed in the experiment. This direct imaging of the Fibonacci modes in a quasi-periodic structure is a clear asset offered by cavity polaritons.

\begin{figure}[htb]
\begin{center}
\includegraphics[width=8.5cm]{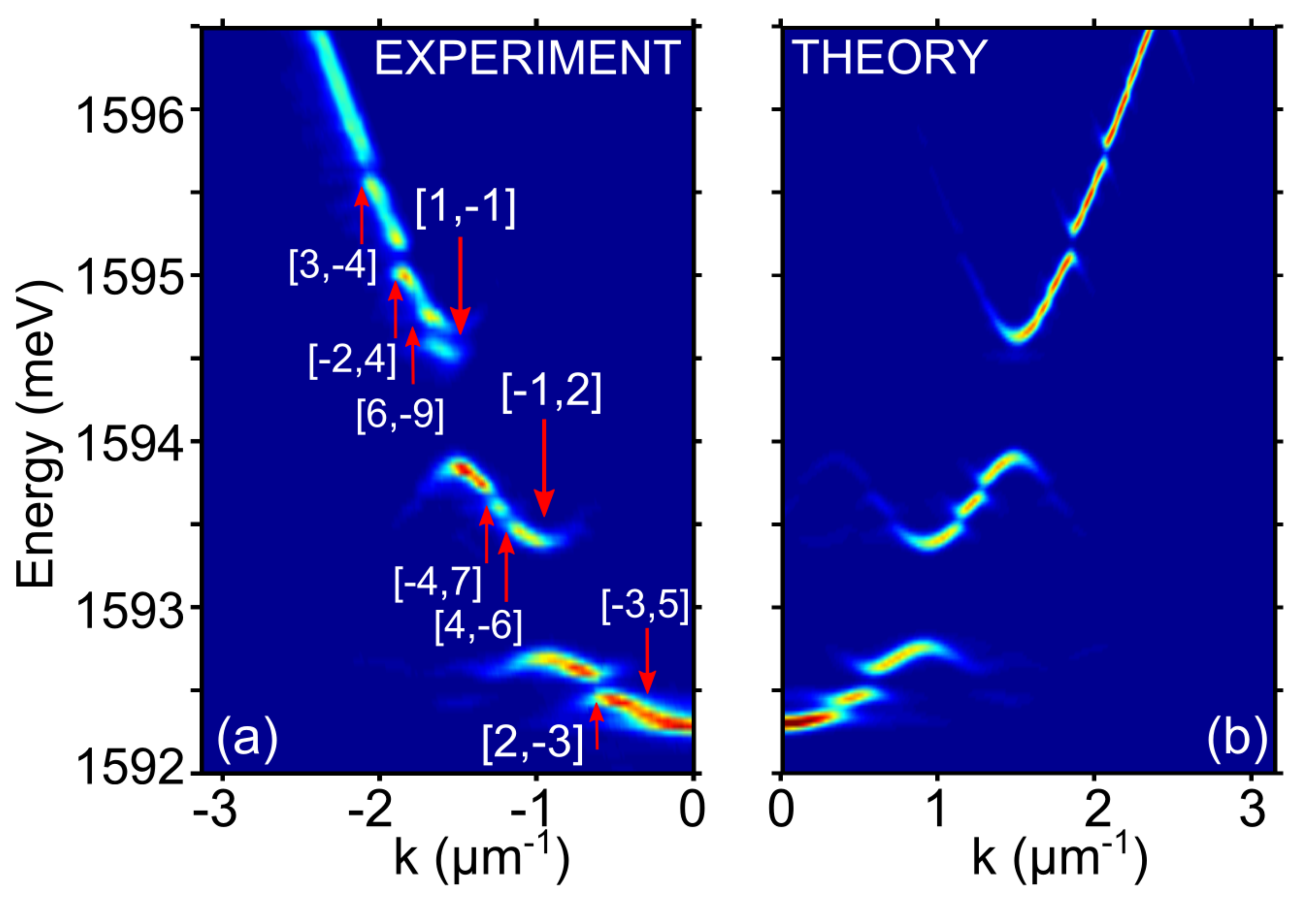}
\end{center}
\caption{\label{fig3}{(Color online) (a) Spectrally resolved far field emission measured on the same wire cavity used in Fig.\ref{fig2}; (b) Corresponding simulation. Position of the gaps labeled with two integers $[p,q]$ is indicated with red arrows.}}
\end{figure}

Probing the polariton modes in reciprocal space provides also remarkable information about the eigenmodes. This is illustrated on Fig.\ref{fig3}.a, where taking advantage of the one-to-one relation between angle of emission and in-plane momentum of polaritons, far field imaging of the polariton emission is shown for the same wire as in Fig.\ref{fig2}. A complex band structure appears with the opening of gaps not regularly spaced unlike the case of a periodic modulation  \cite{Tanese}. The calculated band structure reproduces quantitatively the measurements (Fig.\ref{fig3}.b).

\begin{figure*}[htb]
\begin{center}
\includegraphics[width=15cm]{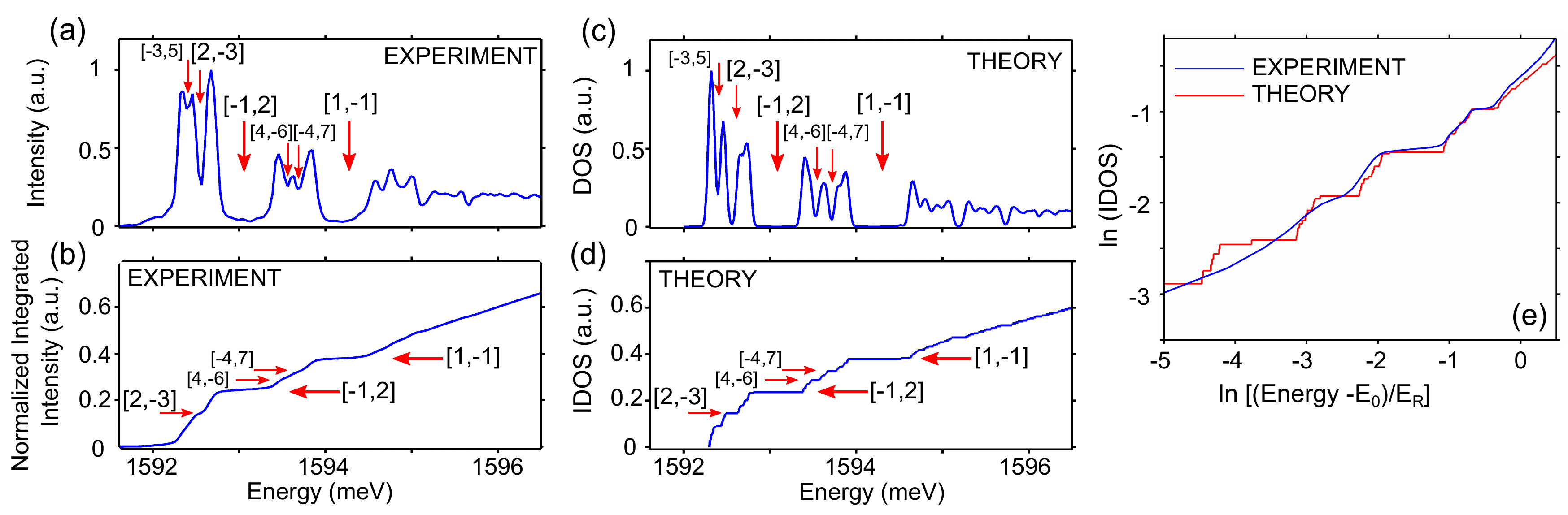}
\end{center}
\caption{\label{fig4} {(Color online) (a) Measured total (angularly integrated) emission spectrum $I(\varepsilon)$ of the quasi-periodic wire and (b) Spectrally integrated emission intensity $\int_{E_0}^{\varepsilon}I(\varepsilon') d\varepsilon'$} (where $E_0$ is the lower energy state); (c-d) Calculated normalized DOS (c) and IDOS (d). (e) Display of the log-periodic oscillations of the IDOS in a $\log$-$\log$ plot of numerical (red) and  experimental (blue) IDOS in the vicinity of $E_0$ (normalized to $E_R=\hbar^2 \pi ^2/(8a^2 m_p)$, with $m_p$ the polariton mass).}
\end{figure*}

In the rest of the paper, we show that despite the finite size of the system, both in the numerics and in the experiments, fundamental physical properties are evidenced  in this complex band-structure which indicate the onset of a fractal density of states.
To study the spectrum and the position of its gaps, it is convenient  to rewrite the quasi-periodic potential $V(x)$ in ({\ref{Effective 1D1}) under the form,
\be
V(x) =  \sum_n \chi (\sigma^{-1}  n) \, u_b (x - a n)
\label{vofx}
\ee
valid in principle \cite{Kohmoto-87} for an infinitely long system namely $j \rightarrow \infty$ in (\ref{Fib Recurs}). $u_b (x)$ (which depends on $w(x)$) describes the shape of the letter $B$ while the periodic  function $\chi (x)$ defined, within $[0,1]$, by $\chi (x) =1$ for $0 <x< 2 -\sigma$ and $\chi (x) =0$ for $ 2 -\sigma <x< 1$, accounts for the quasi-periodic order. The Fourier transform of $V(x)$ consists of Bragg peaks and is given by,
\be
V(k) = \tilde u_b (k) \, \sum_{p,q} \chi_q \, \delta \left(ka-2 \pi (p + q \sigma^{-1} ) \right)
\label{vofk}
\ee
with obvious notations. Since $\sigma$ is irrational, each Bragg peak of the quasi-periodic potential can be uniquely labeled with a set $[p,q]$ of two integers so that the corresponding wave number is
$k = Q_{p,q} \equiv {2 \pi \over a} \left( p + q \sigma^{-1} \right)$. Similarly to the Bloch theorem for a periodic modulation, we may expect that a series of gaps opens at each independent Bragg peak $Q_{p,q}$. Thus, to label the gaps and to obtain the IDOS given in (\ref{Scaling func}), it is tempting to consider the quasi-periodic potential $V(x)$ as a small perturbation. Albeit not justified in the present experimental case, we shall first use this assumption since it allows to give a more intuitive derivation of gap labeling.  But the Bragg peaks being a dense set, we must be cautious and first approximate $\sigma$ by its finite approximants $\sigma_j = F_{j+1} / F_j$ as defined after (\ref{Fib Recurs}). Then, $V(x)$ in (\ref{vofx}) becomes a periodic approximant $V_{j+1} (x)$, built from  periodically repeated cells $S_{j+1}$ of length $a \, F_{j+1}$. Thus, the properties of the single cell  $S_{j+1}$ studied experimentally are essentially those of the periodic potential $V_{j+1} (x)$. Its Fourier transform $V_{j+1} (k)$ is obtained replacing $\sigma$ by $\sigma_j$ in (\ref{vofk}).
$V_{j+1} (k)$ thus defined, is the structure factor of a periodic structure and therefore it has a finite density of Bragg peaks spaced by $\Delta k = 2 \pi / (a F_{j+1} )$. Perturbation theory in $|V| \ll 1$ is now applicable. To first order, each Bragg peak
$k = Q_{p,q} \equiv {2 \pi \over a} \left( F_{j+1} p + F_{j} q \right)$ hybridizes the degenerate Bloch waves at wave numbers $\pm Q_{p,q}/2$. The coupling between these plane waves is best described by a two-level Hamiltonian with diagonal, $\varepsilon \equiv E_{Q_{p,q}/2} = E_{-Q_{p,q}/2}$, and off-diagonal, $V_q \equiv V \chi_q$, matrix elements.
The doubly degenerate level $\varepsilon$ splits into $\varepsilon \pm |V_q |$ and a gap of width $2 |V_q |$ opens at this energy. Accordingly, there is a one-to-one correspondence between the Bragg peaks and the gaps generated through the hybridization of plane waves, so that each gap can also be labeled with the two integers $[p,q]$.
Noting that $Q_{p,q} \, a / 2 \pi = p + q \sigma^{-1}$ is the proportion of unperturbed eigenmodes whose energies are less than $\varepsilon = E_{Q_{p,q}/2}$, the IDOS inside the  $[p,q]$-gap is
\be
\mathcal{N} (\varepsilon = E_{Q_{p,q}/2}) = p + q \sigma^{-1} = q \sigma^{-1} \, \, \mbox{(mod. 1)} \, \,
\label{idosmod1}
\ee
for  $\mathcal{N} (\varepsilon= E_{Q_{p,q}/2}) $ normalized to unity at $E_{Q_{1,0}}$.

While the previous result has been obtained using perturbation theory,  it happens that  it has a much broader range of validity generally expressed by the so called gap labeling theorem \cite{gaplabel} formulated by Bellissard and coworkers. This theorem provides a precise framework for applicability and allows to compute values of the IDOS in the gaps of the spectrum of 1D  Schr\"odinger Hamiltonians with bounded potentials $V(x)$.  An important consequence of that theorem is the topologically stable nature of the IDOS values in the gaps which extends beyond  perturbation theory. Those specific values are obtained \cite{gaplabel} from some prescribed linear combinations of components of eigenvectors of the corresponding substitution matrix characteristic of the quasi-periodic potential. For the Fibonacci sequence defined in (\ref{Fib Recurs}), that prescription reduces to linear combinations of $1$ and $\sigma^{-1}$ namely  to (\ref{idosmod1}).
In Fig.\ref{fig3}.a, we indicate with red arrows the labeling of the gaps using the set $[p,q]$, demonstrating that the positions of the gaps are accurately determined by the positions of the Bragg peaks even for a relatively short Fibonacci sequence such as considered here. These positions are topological quantities, namely independent of the strength of the potential. These observed spectral features are thus independent of the (large enough) sample size and of the realization of the potential. These points are further discussed in the supplement \cite{supplement}. On the other hand, the energy width of the gaps depends  on the heights of the Bragg peaks, {\it i.e.} on the details of the potential $u_b (x)$ (and $w(x)$).

The peculiar structure of the emission spectrum appears also clearly by considering the total  emission intensity $I(\varepsilon)$ nearly proportional to the DOS for low excitation powers. Fig.\ref{fig4}.a displays peaks and deeps corresponding respectively to bands and pseudo-gaps. The measured integrated intensity $ \int_{E_0}^{\varepsilon}I(\varepsilon') d\varepsilon'$ (with $E_0$ being the lower energy state), is reported in Fig.\ref{fig4}.b together with the numerically calculated DOS and IDOS (Figs \ref{fig4}.c-d). Applying (\ref{idosmod1}), valid in principle in the infinite limit, to the gaps $[2,-3],[-1,2],[1,-1]$ indicated in Figs.\ref{fig4}.b-d, gives respectively $\mathcal{N} (E_{Q_{p,q}/2}) =0.15,0.24,0.38$. These numbers are in excellent agreement with the experiment, confirming the good homogeneity achieved in populating the polariton states.

For the infinite system, there exists an infinite series of gaps at $p + q \sigma^{-1} \in [0,1]$ . Thus the energy spectrum, which is the complementary of these gaps, is singular continuous. It is a Cantor like set whose total width vanishes.
The high resolution available in the numerics allows to consider finer details of the IDOS as predicted by the scaling form (\ref{Scaling func}). In Fig.\ref{fig4}.e}, we have plotted in a log-log scale the IDOS as a function of  (properly normalized) energy. It is noticeable that, even for such a finite sized system, we indeed observe a power law behavior multiplied by a log-periodic function. More interesting is the experimental observation of these log-periodic oscillations, showing two periods of oscillations, which constitutes a direct and so far unobserved signature of the fractal character of the Fibonacci spectrum.

In summary, probing the luminescence of a polariton gas laterally confined by a Fibonacci quasi-periodic potential, we have observed the characteristic behavior of the associated fractal energy spectrum: gaps densely distributed, and an IDOS well described by the scaling form (\ref{Scaling func}) and following   the gap labeling theorem (\ref{idosmod1}). We have obtained a spectrally and spatially resolved image of the polariton modes which is in good quantitative agreement with theoretical and numerical results. Our results support the idea that topological features of a fractal spectrum are robust and show up quite accurately even for a relatively short structure. Those results evidence the great interest of cavity polaritons to study the anomalous time expansion of  a polariton wave-packet \cite{Guarneri}, more complex quantum systems {\it e.g.} 2D quasi-crystals \cite{vignolo} and more generally to realize
quantum simulators.

Acknowledgements: This work was supported by the Israel Science Foundation Grant No.924/09, by the 'Agence Nationale pour la Recherche' project "Quandyde" (ANR-11-BS10-001), by the FP7 ITN "Clermont4" (235114) , by the french RENATECH network, the LABEX NanoSaclay and the Honeypol ERC starting grant.

\newpage

\clearpage

\setlength{\topmargin}{-1.35in}
\setlength{\oddsidemargin}{-1.1in}
\includegraphics[page={1},width=1.165\textwidth]{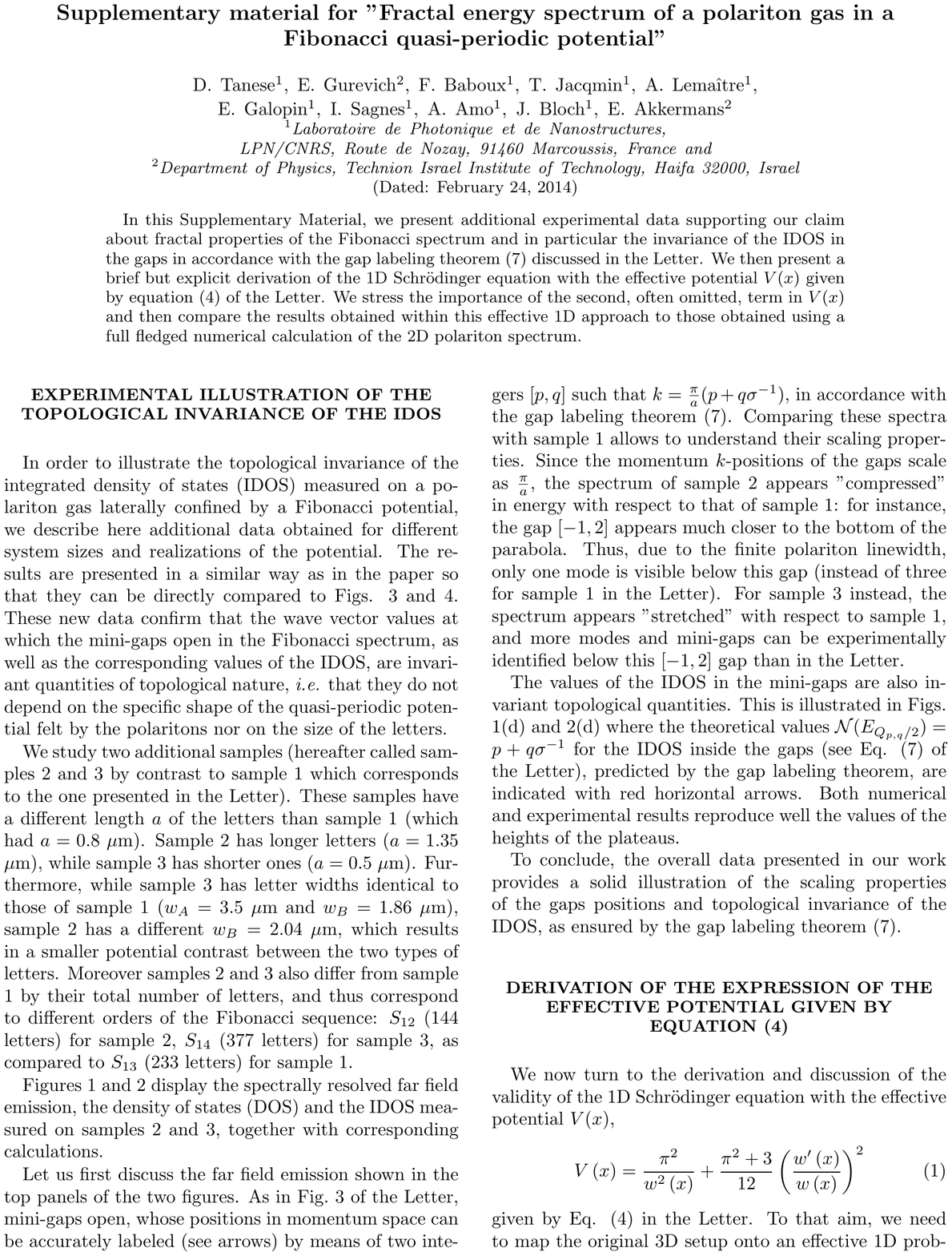}
\clearpage
\setlength{\oddsidemargin}{-1.1in}
\includegraphics[page={2},width=1.165\textwidth]{SuppMat.pdf}
\clearpage
\setlength{\oddsidemargin}{-1.1in}
\includegraphics[page={3},width=1.165\textwidth]{SuppMat.pdf}
\clearpage
\setlength{\oddsidemargin}{-1.1in}
\includegraphics[page={4},width=1.165\textwidth]{SuppMat.pdf}
\clearpage
\setlength{\oddsidemargin}{-1.1in}
\includegraphics[page={5},width=1.165\textwidth]{SuppMat.pdf}


\begin{thebibliography}{99}
\bibitem{singcontgeneral} H.L. Cycon, R.G. Froese, W. Kitsch and B. Simon, Schr\"odinger Operators, (Springer, Berlin, 1987) and M. Reed and B. Simon, Methods of Modern Mathematical Physics (Academic Press, California,
1980).

\bibitem {damanik}D. Damanik and A. Gorodetski, Commun. Math. Phys.
\textbf{305}, 221 (2011) and D. Damanik, M. Embree, A. Gorodetski, S. Tcheremchantsev,  Commun. Math. Phys. \textbf{280}, 499 (2008)

\bibitem{vardeny} For a recent review see Z. V. Vardeny, A. Nahat and  A. Agrawal, Nature Photonics 7, 177ðËü187 (2013).


\bibitem {Kohmoto-87}
M. Kohmoto, B. Sutherland and C. Tang, Phys. Rev. B \textbf{35},
1020 (1987); J.M. Luck, Phys. Rev. B \textbf{39}, 5834 (1989).

\bibitem{gelerman} W. Gellermann, M. Kohmoto, B. Sutherland and P.C. Taylor , Phys. Rev. Lett. {\bf 72}, 633  (1994).



\bibitem {QP-Review03}M. Kohmoto, L.P. Kadanoff and C. Tang, Phys. Rev.\ Lett.
\textbf{50}, 1870 (1983) and S. Ostlund and S.Kim, Physica Scripta \textbf{9},
193 (1985). For a review see E.L. Albuquerque and M.G. Cottam, Phys. Rep.
\textbf{376}, 225 (2003); E. Maci\'{a}, Rep. Prog. Phys. \textbf{69}, 397 (2006).

\bibitem {Kohmoto-87-Wurtz-88}M. Kohmoto, B. Sutherland and K. Iguchi, Phys.
Rev. Lett. \textbf{58}, 2436 (1987) ; D. W\"{u}rtz, T. Schneider and M.P.
Soerensen, Physica A \textbf{148}, 343 (1988).



\bibitem{reviewfractals} For a recent review, E. Akkermans, Contemporary Mathematics {\bf 601}, 1-22  (2013), arXiv:1210.6763.


\bibitem {ADT2}E. Akkermans, G.V. Dunne and A. Teplyaev, Phys. Rev. Lett.
\textbf{105}, 230407 (2010).


\bibitem{ADT3} E. Akkermans, O. Benichou, G. Dunne, A. Teplyaev and  R. Voituriez,
Phys. Rev. E {\bf 86}, 061125 (2012).

\bibitem{Guarneri} I. Guarneri and G. Mantica, Phys. Rev. Lett. \textbf{73}, 3379 (1994) and S. Abe and H.  Hiramoto, Phys. Rev. A {\bf 36}, 5349 (1987).

\bibitem{eagurevich} E.~Akkermans and E.~Gurevich, Europhys. Lett.  {\bf 103}, 30009 (2013).
\bibitem{weisbuch92} C. Weisbuch, M. Nishioka, A. Ishikawa, and Y. Arakawa, Phys. Rev. Lett. \textbf{69}, 3314 (1992)
\bibitem{LaiNature2007} C. W. Lai, N. Y. Kim, S. Utsunomiya, G. Roumpos, H. Deng, M. D. Fraser, T. Byrnes, P. Recher, N. Kumada, T. Fujisawa and Y. Yamamoto, Nature \textbf{450}, 529 (2007)
\bibitem{CarusottoRMP} I. Carusotto and C. Ciuti, Rev. Mod. Phys. \textbf{85}, 299 (2013)
\bibitem{CerdaPRL}E. A. Cerda-M\'endez, D. N. Krizhanovskii, M. Wouters, R. Bradley, K. Biermann, K. Guda, R. Hey, P. V. Santos, D. Sarkar, and M. S. Skolnick, Physical Review Letters \textbf{105}, 116402 (2010)
\bibitem{Tanese} D. Tanese, H. Flayac, D. Solnyshkov, A. Amo, A. Lema\^{\i}tre, E.Galopin, R. Braive, P. Senellart, I. Sagnes, G. Malpuech and J. Bloch, Nature Communication \textbf{4}, 1749 (2013)
\bibitem{NaYongNaturePhysics2011}   N. Y. Kim, K. Kusudo,C. Wu, N. Masumoto, A. L\"offler,S. H\"ofling, N. Kumada, L. Worschech, A.Forchel and Y. Yamamoto, Nature Physics \textbf{7},681 (2011)
\bibitem{CerdaDots}E. A. Cerda-M\'endez, D. N. Krizhanovskii, K. Biermann, R. Hey, M. S. Skolnick, and P. V. Santos, Phys. Rev. B \textbf{86}, 100301 (2012).
\bibitem{Cerda Gap solitons}E. A. Cerda-M\'endez, D. Sarkar, D. N. Krizhanovskii, S. S. Gavrilov, K. Biermann, M. S. Skolnick, and P. V. Santos, Phys. Rev. Lett. \textbf{111}, 146401 (2013)
\bibitem{Nguyen} H. S. Nguyen, D. Vishnevsky, C. Sturm, D. Tanese, D. Solnyshkov, E. Galopin, A. Lema\^{\i}tre, I. Sagnes, A. Amo, G. Malpuech, and J. Bloch
Phys. Rev. Lett. \textbf{110}, 236601 (2013)
\bibitem{NaYongNJPhysics} N. Y. Kim, K. Kusudo, A. L\"offler, S. H\"ofling, A. Forchel,and Y. Yamamoto, New Journal of Physics \textbf{15}, 035032 (2013)
\bibitem{yamamotoHoneycomb}  N. Y. Kim, A. L\"offler, S. H\"ofling, A. Forchel, and Y. Yamamoto, Phys. Rev. B \textbf{87}, 214503 (2013)
\bibitem{TJacqmin} T. Jacqmin, I. Carusotto, I. Sagnes, M. Abbarchi, D. Solnyshkov, G. Malpuech, E. Galopin, A. Lema\^{\i}tre, J. Bloch and A. Amo, arXiv:1310.8105 (2013).
\bibitem{supplement} See supplementary materials.
\bibitem{dalnegro} L. D. Negro, C. J. Oton, Z. Gaburro, L. Pavesi, P. Johnson, A. Lagendijk, R. Righini, M. Colocci, and D. S. Wiersma, Phys. Rev. Lett. \textbf{90}, 055501 (2003).

\bibitem{gaplabel} J. Bellissard, A. Bovier and J.M. Ghez, Reviews in Math. Physics, Vol. 4, No. 1, 1-37 (1992) and B. Simon, Adv. Appli. Math. {\bf 3}, 463 (1982) and J. Bellissard, Les Houches, Springer, J.M. Luck, P. Moussa
and M. Waldschmidt Eds., (1993).

\bibitem{vignolo}  J-M. Gambaudo and  P. Vignolo, arXiv:1309.6420 (2013).


\end{thebibliography}
\end{document}